\documentclass[a4paper,twocolumn,superscriptaddress,nofootinbib]{revtex4-1}
\usepackage{hyperref}
\usepackage[T1]{fontenc}
\usepackage[utf8]{inputenc}
\usepackage{lmodern}
\usepackage{amsmath}
\usepackage{amsfonts}
\usepackage{amsthm}
\usepackage{bbold}
\usepackage[retainorgcmds]{IEEEtrantools}
\usepackage{graphicx}
\usepackage{tikz}
\usepackage[justification=RaggedRight]{caption}
\usepackage{subfigure}
\usepackage{color}
\usepackage{booktabs}
\usepackage{multirow}
\usepackage{pifont}
\usepackage{tabularx}

\newcommand{\be}{\begin{equation}}
\newcommand{\ee}{\end{equation}}
\newcommand{\ba}{\begin{eqnarray}}
\newcommand{\ea}{\end{eqnarray}}
\newcommand{\ban}{\begin{eqnarray*}}
\newcommand{\ean}{\end{eqnarray*}}
\newcommand{\nthm}[2]{\newtheorem}

\newcommand{\ket}[1]{\mbox{$ | #1 \rangle $}}

\newcommand{\ketbra}[2]{|#1\rangle\kern-2.8pt\langle#2|}

\newcommand{\norm}[1]{\left|#1\right|}

\newcommand{\tr}{\text{tr}}

\definecolor{nred}{rgb}{0.7,0.2,0.2}
\definecolor{nblack}{rgb}{0,0,0}
\definecolor{nblue}{rgb}{0.2,0.2,0.8}
\definecolor{ngreen}{rgb}{0.2,0.6,0.2}

\usepackage{paralist}

\begin{document}

\renewcommand{\today}{\number\day\space\ifcase\month\or
   January\or February\or March\or April\or May\or June\or
   July\or August\or September\or October\or November\or December\fi
   \space\number\year}

\begin{center}
\title{Nonlocality of $W$ and Dicke states subject to losses}

\date{\today}

\author{Tomer~Jack~Barnea}
\affiliation{Group of Applied Physics, University of Geneva, 1211 Geneva 4, Switzerland}
\author{Gilles~P\"utz}
\affiliation{Group of Applied Physics, University of Geneva, 1211 Geneva 4, Switzerland} 
\author{Jonatan~Bohr~Brask}
\affiliation{Département de Physique Théorique, Université de Genève, 1211 Geneva 4, Switzerland}
\author{Nicolas~Brunner}
\affiliation{Département de Physique Théorique, Université de Genève, 1211 Geneva 4, Switzerland} \author{Nicolas~Gisin}
\affiliation{Group of Applied Physics, University of Geneva, 1211 Geneva 4, Switzerland}
\author{Yeong-Cherng~Liang}
\affiliation{Institute for Theoretical Physics, ETH Zürich, 8093 Zurich, Switzerland}

\begin{abstract}
We discuss the nonlocality of the $W$ and the Dicke states subject to losses. We consider two noise models, namely loss of excitations and loss of particles, and investigate how much loss can be tolerated such that the final state remains nonlocal. This leads to a measure of robustness of the nonlocality of Dicke states, with a clear physical interpretation. Our results suggest that the relation between nonlocality and entanglement of Dicke states is not monotonous.
\end{abstract}

\maketitle

\end{center}

\section{Introduction}

Quantum nonlocality \cite{bell}---the fact that quantum statistics can lead to Bell inequality violations---is now considered a fundamental aspect of quantum theory, and represents a powerful resource for information processing, see e.g. \cite{acin07,review14}. While quantum nonlocality has been extensively studied in the case of two parties \cite{review14}, the multipartite case is not as well understood. This is partly due to the complexity of multipartite entanglement \cite{horodecki09,guehne09} and to the lack of tools adapted to the study of the multipartite nonlocality (see, however, Ref.~\cite{bancal12,liang14}).

In the present paper, we discuss the nonlocal properties of an important class of multipartite entangled states, namely (symmetric qubit) Dicke states \cite{dicke}. These are central in the fields of quantum optics and quantum information processing, as they play a crucial role in the theory of interaction of light and matter \cite{dicke}, in quantum memories \cite{wolfgang09} and are relevant for quantum metrology \cite{lucke11,geza12,hyllus}. Dicke states form a basis of all symmetric multipartite qubit states, and their entanglement properties have been discussed, e.g. in Refs. \cite{andrew03,geza07,marcus1,duan1,guhne13}. 

It is a well known fact that all multipartite entangled pure states violate a Bell inequality \cite{popescu} (see also \cite{yu12}), hence all Dicke states exhibit nonlocality. Moreover, the nonlocality of symmetric pure qubit states is elegantly captured by a single Bell inequality \cite{wang12}. The nonlocality of the simplest Dicke states, featuring a single excitation (the so-called $W$-states), has been widely discussed \cite{cabello,sen,brunner,wang13,bancal11,moroder13,bancal13}, in particular in the context of optical Bell tests based on single photon entanglement \cite{heaney,brask1,brask2}. Notably, the possibility of self-testing the $W$ state has been recently demonstrated \cite{wu,miguel}. Finally, the relevance of the nonlocality of Dicke states in the context of many-body physics has been recently discussed \cite{jordi}.

Our main focus here is to determine the robustness of the nonlocality exhibited by Dicke states with respect to loss. This provides a natural way to quantify the nonlocality of these states, with a clear physical meaning. In addition, this allows us to compare different Dicke states from the point of view of nonlocality. For instance, a basic question is the following: for a given number of particles (or modes) $n$, what is the most robust Dicke state, i.e. how many excitations $k$ are optimal in terms of loss-resistance?

Specifically, we consider two models of losses: (i) loss of excitations, and (ii) loss of particles. For a given Dicke state, our goal is to determine how much loss can be tolerated such that the final state remains nonlocal, i.e. still violates a Bell inequality \cite{persistency,laskowski10}. Our focus is to derive bounds for the case of Dicke states featuring a large number of particles or modes. Moreover, we study how the robustness is influenced by the number of excitations in the state. While the most entangled Dicke state of $n$ particles is the one with exactly $k=\lfloor n/2\rfloor$ excitations \cite{geza07}, we find a very different behavior for nonlocality. Specifically, the most robust Dicke state seems to feature only few excitations, for both types of losses. This suggests that the entanglement and the nonlocality of Dicke states might be non-monotonously related. Note that in the bipartite case, entanglement and nonlocality were proven to behave very differently in certain situations, quite different, however, from the ones studied here, see \cite{review14}.

\section{Scenario}

We consider a source producing a symmetric (qubit) Dicke state
\ba  \label{nk}
   \ket{n,k} & = & {n\choose k}^{-\frac{1}{2}} \text{sym}\left[\ket{0}^{\otimes n-k}\ket{1}^{\otimes k}\right]
\ea 
where $ \text{sym}\left[\ldots\right]$ denotes symmetrization by party-exchange. We refer to such a state as a Dicke state with $n$ particles (or modes) and $k$ excitations. The case $k=1$ corresponds to the so-called $W$ state~\cite{wstate}. We also write \mbox{$\rho_{n,k} = \ketbra{n,k}{n,k}$}. Note that $\rho_{n,0}$ corresponds to the $n$-partite vacuum.

After being emitted by the source, the state $\ket{n,k}$ may undergo some losses, e.g. via propagation through a lossy channel. In the end, local measurements are performed on the final state, and our goal is to characterize the robustness of the nonlocality of the original state with respect to losses, and hence the nonlocal property of this final state. Specifically, we consider two different loss models, the study of which we briefly motivate from a physical point of view.

For the first model, we consider $\ket{n,k}$ as describing the state of a system with $n$ modes featuring $k$ excitations. For instance this could represent $k$ photons distributed among $n$ modes, with an optical loss $p$ in each arm, or alternatively, $k$ excitations stored in an ensemble of $n$ atoms with a decay from state $\ket{1}$ to $\ket{0}$ with probability $p$, e.g. due to spontaneous emission or collisions. In this case, it is natural to discuss channel losses in the following way. In each mode, an excitation has a probability $p$ of being lost. That is, the channel we consider implements the local, but non-unitary amplitude-damping transformation $T$ characterized by the following relations: $\ket{1} \rightarrow \ket{0}$ with probability $p$, otherwise we have $\ket{1} \rightarrow \ket{1}$, while the vacuum component always remains unchanged, i.e. $\ket{0} \rightarrow \ket{0}$ with probability one. Hence the final state is given by 
\ba \rho_f  = T (\rho_{n,k} ). \ea
We refer to this case as `losing excitations', and our main goal is to determine how much loss can be tolerated, i.e. how large $p$ can be such that the final state $\rho_f$ is still nonlocal.

In the second loss model, we view the state $\ket{n,k}$ as that of a system with $n$ particles, among which $k$ are in state $\ket{1}$ whereas the remaining ones are in state $\ket{0}$, where $\ket{0}$ and $\ket{1}$ refer to an internal degree of freedom of each particle. Consider for instance the loss of particles from an atomic ensemble. Here we discuss the case in which a given number of particles $m$ is lost. Hence the final state is given by
\ba \label{lp}
   \tau_f  =  \text{tr}_{m}(\rho_{n,k})
\ea
where $\text{tr}_{m}$ means the partial trace over $m$ fixed particles. Note that since the state $\rho_{n,k}$ is symmetrical, it does not matter which particles are lost. The final state $\tau_f$ contains $n_f = n-m$ particles. We refer to this case as `losing particles', and our objective is to find out the largest fraction of particles that can be lost such that the final state $\tau_f$ remains nonlocal. 

The state after losses is distributed between $N$ observers. Note that $N=n$ for the case of losing excitations, while $N= n_f$ for the case of losing particles. Each observer performs one out of two possible local measurements on his mode or particle. Here we assume that all observers perform the same projective qubit measurements described by the operators
\ba \label{measparam} \mathcal{A}_j = \cos(\alpha_j) \sigma_z +  \sin(\alpha_j) \sigma_x \ea 
where $\sigma_{x,z}$ denote the usual Pauli matrices, each $\alpha_j$ is a real number and $j=0,1$ denotes the choice of setting. It is worth commenting on this choice of measurements. First, given that the final state is a mixture of Dicke states, a rather natural computational simplification is to adopt the same measurement settings for all parties. Second, since the correlations of Dicke states are invariant under the exchange of $x$ and $y$, we chose to focus on settings in the $x-z$ plane of the Bloch sphere.

The resulting measurement statistics are given by joint conditional probabilities
\ba   P(a_1 \dots a_N|x_1 \dots x_N) =  \tr( \rho \mathcal{P}^{a_1}_{x_1} \otimes \dots \otimes \mathcal{P}^{a_N}_{x_N} ) \ea
where $x_i=0,1$ and $a_i=\pm1$ denote the measurement choice and outcome, respectively, for observer $i$. Note that we have used the projectors $\mathcal{P}^{a_i}_{x_i} = (\mathbb{I} + a_i \mathcal{A}_{x_i})/2$ here. In order to test the nonlocality of this correlation, we restrict ourselves to a Bell scenario with two binary-outcome measurements per observer. We shall make use of two specific Bell inequalities which have generalizations for $N$ parties. The first is given by
\ba \mathcal{S}_N &=& P(0\dots 0|0 \dots 0) - \sum_{\pi} P(0\dots 0|\pi(0 \dots 0 1)) \nonumber\\
\label{SN}& & - P(1\dots 1|1 \dots 1) \leq 0, \ea
where the sum goes over all $N$ permutations of $(0\cdots 0 1)$. This inequality (first discussed in \cite{larsson01}, see also \cite{wang12}) can be viewed as a multipartite generalization of the Hardy paradox \cite{hardy93}. The second is the full correlation Bell inequality of Mermin-Ardehali-Belinskii-Klyshko (MABK) \cite{mermin90,ardehali92,belinskii93},
\ba\label{MN} \mathcal{M}_N = \norm{\sum_{\vec{x} \in \{0,1\}^{\otimes N}} \beta(x,N) E(\vec{x})} \leq 2^N, \ea
where $\vec{x}=(x_1\cdots x_N)$ is the vector of all inputs, $x = \sum_{k=1}^N x_k$, \ba E(\vec{x})=\sum_{a_1\cdots a_N}\left(\prod_i{a_i}\right)P(a_1\cdots a_n|x_1 \dots x_N)\ea
and~\cite{Wallman:PRA:2011} 
\ba \beta(x,N)=2^\frac{N+1}{2} \cos\left[\frac{\pi}{4} (1+N-2x)\right].\ea
Note that the Hardy Bell expression \eqref{SN} involves a number of joint probabilities that grows linearly with $N$, while the MABK Bell expression \eqref{MN} features a number of correlation functions that grows exponentially with $N$, which renders the numerical analysis of a large $N$ more managable for the former.

We denote by $\mathcal{S}_N(\rho,\alpha_0,\alpha_1)$ and $\mathcal{M}_N(\rho,\alpha_0,\alpha_1)$ the values that are obtained for $\mathcal{S}_N$ and $\mathcal{M}_N$ by performing the measurements parametrized by the measurement angles $\alpha_0$ and $\alpha_1$ [cf. Eq.~\eqref{measparam}] on the state $\rho$. In the following we use these two quantities to characterize the nonlocality of different Dicke states, starting with the $W$ state, after they have been subjected to the two different types of losses.

\section{$W$ state}
\label{secW}

We start our investigation with the single-excitation Dicke state (i.e. $k=1$), also known as the $W$ state. 
For our first model of losses, i.e. losing excitations with probability $p$, the final state is given by
\ba \label{leW} \rho_f =  (1-p) \rho_{n,1} + p \rho_{n,0}.  \ea
For our second model, i.e. losing particles, the final state is given by [cf Eq.~\eqref{lp}]
\ba  \label{lpW} \tau_f = \frac{n_f}{n} \rho_{n_f,1} + \frac{n-n_f}{n} \rho_{n_f,0}. \ea
Although the number of modes is different in the two cases, the problem of determining the robustness of the final state is essentially the same. It boils down to finding the robustness of the nonlocality of the pure $W$ state with respect to mixing with the (separable) state $\rho_{N,0}$. Note that the state $\rho_f$ is entangled for any $p<1$. Determining the robustness of the nonlocality of the $W$ state (for general $n$) with respect to losing excitations also determines the robustness with respect to the other loss model. If we assume that a probability $p$ of losing excitations can be tolerated for $N$ parties, then this implies that for $n=\lceil \frac{N}{1-p}\rceil$ parties, $n-N$ particles can be lost while preserving the nonlocality of the state. It is therefore sufficient to study the first model.

To this end, we now focus on the $n$-dependency of the maximal loss probability $p$, denoted by $p_{th}(n)$, such that $\rho_f$ is nonlocal for all $p<p_{th}(n)$. By analyzing the violation of $\mathcal{S}_n$ and $\mathcal{M}_n$ using the simplification given in Eq.\eqref{measparam}, we get lower bounds on $p_{th}(n)$, which we denote by $p^\mathcal{S}_{th}(n)$ and $p^\mathcal{M}_{th}(n)$ respectively. In the case of the $W$ state, these lower bounds can be found by performing the optimizations
\ba \label{pS}p^{\mathcal{S}}_{th}(n)=\max_{\alpha_0,\alpha_1}\frac{\mathcal{S}_n(\rho_{n,1},\alpha_0,\alpha_1)}{\mathcal{S}_n(\rho_{n,1},\alpha_0,\alpha_1)-\mathcal{S}_n(\rho_{n,0},\alpha_0,\alpha_1)}\ea
and
\ba \label{pM}p^{\mathcal{M}}_{th}(n)=\max_{\alpha_0,\alpha_1}\frac{\pm 2^n-\mathcal{M}_n(\rho_{n,1},\alpha_0,\alpha_1)}{\mathcal{M}_n(\rho_{n,0},\alpha_0,\alpha_1)-\mathcal{M}_n(\rho_{n,1},\alpha_0,\alpha_1)}.\ea

To perform this optimization, we computed $\mathcal{S}_n(\rho_{n,1},\alpha_0,\alpha_1)$ and $\mathcal{S}_n(\rho_{n,0},\alpha_0,\alpha_1)$ and the corresponding terms for $\mathcal{M}_n$ as a function of $\alpha_0$ and $\alpha_1$ (see Appendix \ref{A}). For small $n$, the optimization can be carried out for both cases. To extend the result to large $n$, for which the numerical optimization becomes computationally infeasible, we used the optimal measurement angles for small $n$ to guess their dependency on $n$. The resulting ansatz that we adopted for the Hardy inequality $\mathcal{S}_n$ is given by
\ba \alpha^{\mathcal{S}}_{0}\left(n\right) & = & \frac{\pi}{2} - \arctan\left(\sqrt{7n}\right), \nonumber \\
\alpha^{\mathcal{S}}_{1}\left(n\right) & = & 1 - \frac{1}{\pi}\arctan\left(\sqrt{12n}\right).\ea
For the MABK inequality $\mathcal{M}_n$ it turns out that we need to differentiate the four cases of $n=0,1,2,3 \text{ mod } 4$. The corresponding functions for $\alpha_j^{\mathcal{M}}$ can be found in Appendix \ref{B}.

\begin{figure}[htb]
\begin{center}
\includegraphics[width =\columnwidth]{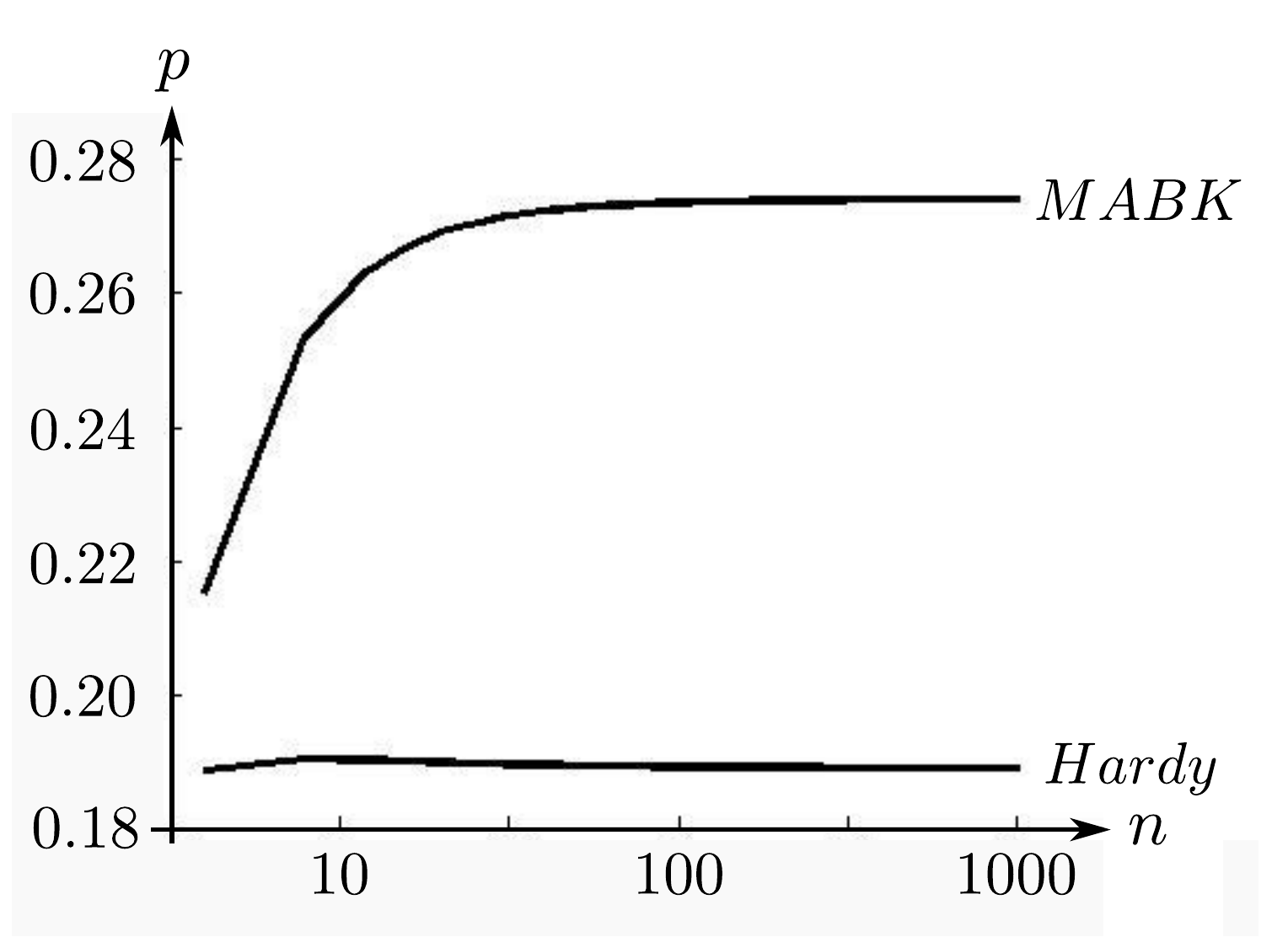}
\caption{For $W$ states of $n$ qubits, the graph shows lower bounds on the threshold probability $p_{th}(n)$ of losing an excitation for the Hardy and MABK inequalities. The best bound is obtained using the MABK inequality, $p_{th}(n) \geq 27.41\%$ for large $n$. For the Hardy inequalitie we get $p_{th}(n) \geq 18.89\%$ for large $n$.}
\label{Wgraph}
\end{center}
\end{figure}

Computing Eq.~\eqref{pS} and Eq.~\eqref{pM} for these angles gives us lower bounds on $p^{\mathcal{S}}_{th}(n)$ and $p^{\mathcal{M}}_{th}(n)$ which are shown in Fig.~\ref{Wgraph}. It can be seen that the bounds provided by the MABK inequality, for which results could be achieved only up to $n= 10^3$ for computational reasons, are better than those given by the Hardy inequality. However, for the Hardy inequality we can determine the asymptotic behavior ($n \to \infty$), yielding $p^{\mathcal{S}}_{th}(n\rightarrow \infty) \geq 18.89\%$, we were not able to do the same for the MABK inequality. Given that the difference between the bounds provided by the two inequalities remains very large, we conjecture nonetheless that the MABK inequality is better suited to the task for all $n$ and that $p_{th}(n)\geq p^{\mathcal{M}}_{th}(n=10^3) \geq 27.41\%$ for large $n$. Finally, in the context of losing particles the above results show that at least a constant fraction of $p$ can be lost for large $n$.

The lower bounds for $p_{th}(n \rightarrow \infty)$ have to be compared to the results of Refs.~\cite{laskowski10,brask1}, where it was bounded by $\frac{1}{3}$.

\section{Dicke states}
\label{secDicke}

We now turn our attention towards general Dicke states with $k$ excitations. Let us first note that, unlike the case of the $W$ state, here the two loss models have to be treated seperately. This can be seen already for the case of $k=2$ excitations, for which we find
\ba \rho_f=p^2 \rho_{n,2} + 2p(1-p)\rho_{n,1} + (1-p)^2\rho_{n,0} \ea
and
\ba \tau_f = \sum_{l=0,1,2} \frac{\binom{n_f}{l} \binom{n-n_f}{2-l}}{\binom{n}{2}} \rho_{n_f,l} \ea
We therefore have to treat the two loss models separately. Note also that the Dicke states with $k$ and $n-k$ excitations are equivalent when exchanging the role of $\ket{0}$ and $\ket{1}$. This symmetry is preserved in the final state $\tau_f$ after particle loss, but not in the final state $\rho_f$ after losing excitations. This is due to the fact that the loss model of losing excitations introduces an asymmetry between $\ket{0}$ and $\ket{1}$. We therefore conclude that the same number of particles can be lost for Dicke states with $k$ or $n-k$ excitations. On the contrary this is not the case for losing excitations, where the amplitude-damping channel has a different effect on the states $\ket{0}$ and $\ket{1}$.

\subsection{Losing excitations}
We start our analysis by looking at the generalized Hardy inequality $\mathcal{S}_n$. The expressions for $\mathcal{S}_n(\rho_{n,k},\alpha_0,\alpha_1)$ can be found in Appendix \ref{C}. As in the case of the $W$ state, we performed the optimization over $\alpha_0$ and $\alpha_1$ in the case of few parties for different numbers of excitations $k=2\ldots 6$ and investigated the dependence of the optimal measurement angles on the number of parties $n$. For $k=2$, the resulting ansatz is given by
\ba\alpha_0^{\mathcal{S}}(n,2) &=& \frac{\pi}{2} - \arctan(1.97 \sqrt{n}) \nonumber \\
    \alpha_1^{\mathcal{S}}(n,2) &=&\frac{\pi+3}{2} - \arctan(6.93 \sqrt{n}).\ea
For $k=3\ldots 6$, we found that we could choose functions which are structurally the same and given by
\ba\alpha^{\mathcal{S}}_{0}(n,k) & = &\frac{\pi}{2} - \arctan(q_{0}(k)\sqrt{n}) \nonumber \\
\label{alpha36Hardy}\alpha^{\mathcal{S}}_{1}(n,k) & = &\frac{\pi+1}{2} - \arctan(q_{1}(k)\sqrt{n}).\ea
The values of the coefficients $q_{j}(k)$ as well as the lower bounds on the threshold probability given by these measurements can be found in Table \ref{pthDicke}.

\begin{table}[b]
  \begin{tabular}{||l|c|c|c||}
  \hline
  \hline
	$k$ & Lower bound on $p^{\mathcal{S}}_{th}(n=10^4)$ & $q_{0}(k)$ & $q_{1}(k)$\\
	\hline
	1 & 0.1889 & & \\ 
	2 & 0.2599 & & \\
	3 & 0.2837 & 1.63 & 4.72 \\
	4 & 0.2956 & 1.47 & 3.77 \\
	5 & 0.2994 & 1.34 & 3.07 \\
	6 & 0.3017 & 1.24 & 2.66 \\
  \hline
  \hline
\end{tabular}
\caption{For Dicke states with $n=10^4$ and $k$ excitations, we give lower bounds on the threshold probability of losing excitations considering the Hardy inequality. Also given are the values of the coefficients $q_{0}(k)$ and $q_{1}(k)$ defining the measurement angles given by Eq.~\eqref{alpha36Hardy}. Note that these coefficients were found with respect to the optimal angles for small values of $n$.}
\label{pthDicke}
\end{table}

\begin{figure}[t]
\begin{center}
\includegraphics[width = \columnwidth]{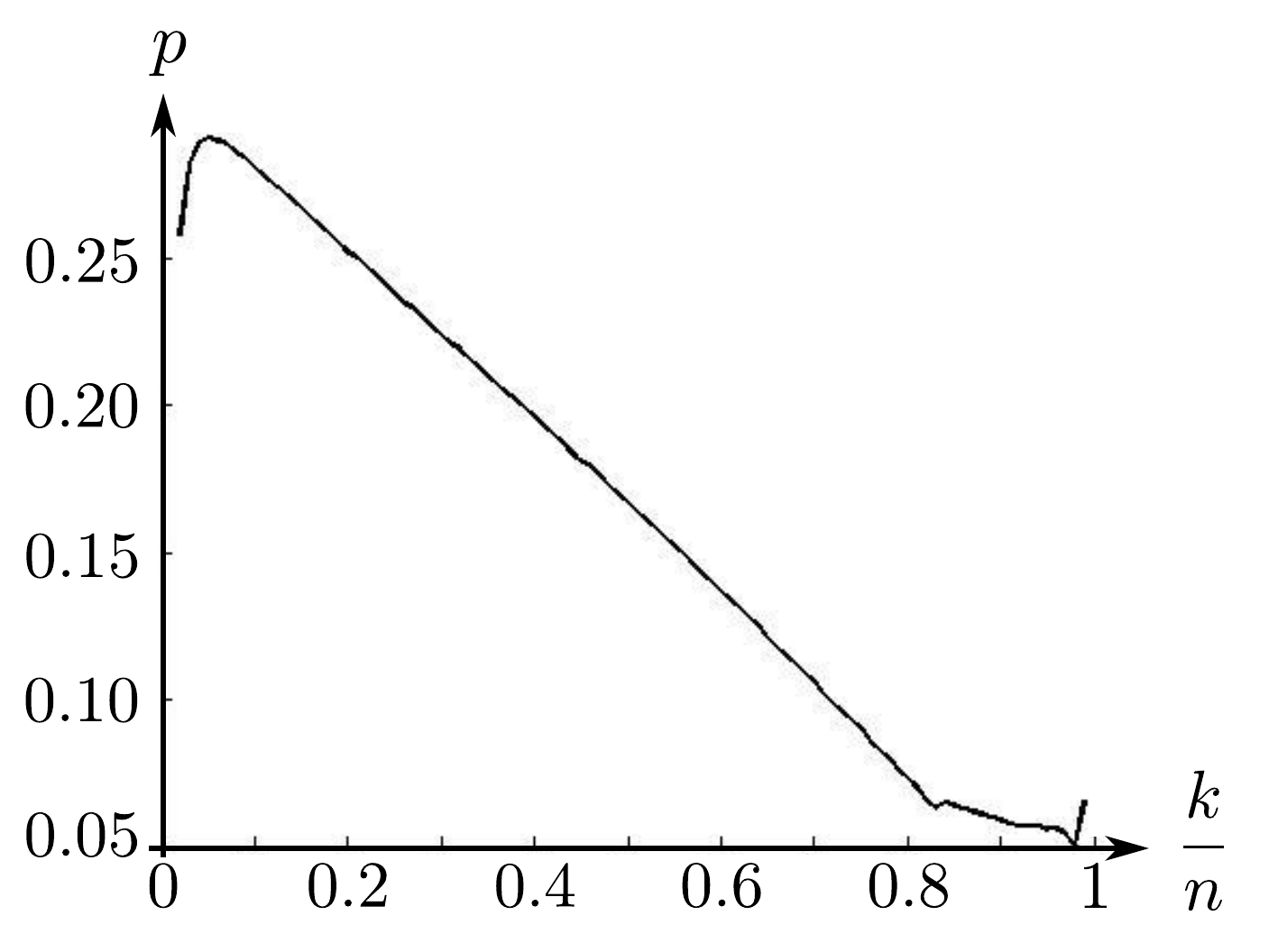}
\caption{Bounds on the threshold probability $p_{th}(100,k)$ of losing an excitation for Dicke states of $n=100$ qubits with $k$ excitations. Here we consider the Hardy inequality $\mathcal{S}_{100}$. The most robust Dicke state features only few excitations, here $k = 5$. Note that for $\frac{k}{n}$ approaching $1$ the numerics become unstable.}
\label{HardyEx100}
\end{center}
\end{figure}

\begin{figure}[htb]
\begin{center}
\includegraphics[width = \columnwidth]{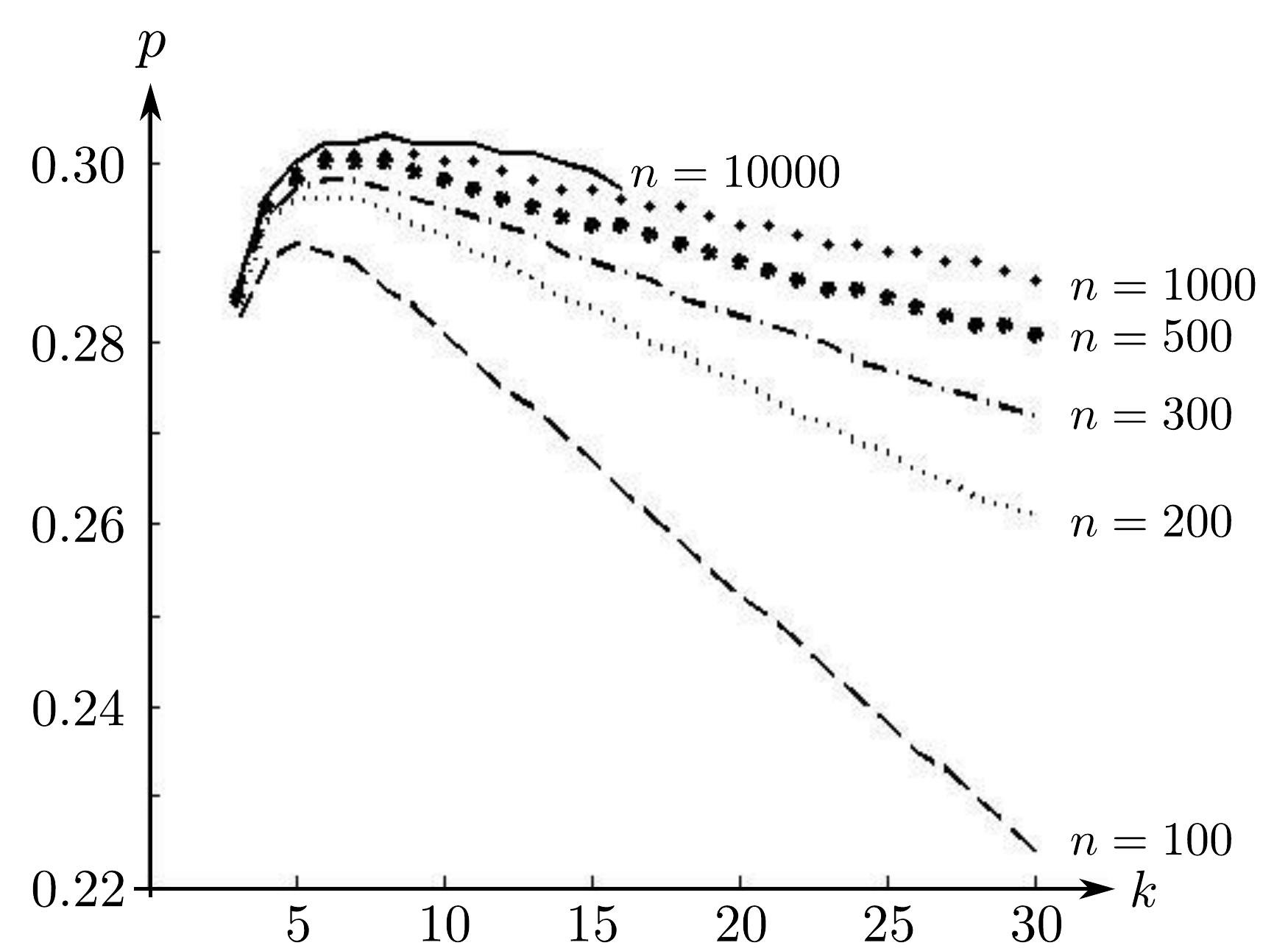}
\caption{We consider Dicke states of $n$ qubits with $k$ excitations, and give bounds $p_{th}(n,k)$. The bottommost curve, i.e. $n=100$, corresponds to Fig. \ref{HardyEx100}. Clearly, robustness is increased as the number of qubits $n$ increases. For large values of $n$, numerics could only be performed for small $k$.}
\label{HardyExN}
\end{center}
\end{figure}

When looking at Table \ref{pthDicke}, it may seem that the threshold probability always increases monotonically with the number of excitations, as one may expect due to the fact that the state with $k=\lfloor\frac{n}{2}\rfloor$ excitations contains the largest amount of entanglement. This is however not supported by the results that we obtained. For fixed values of $n$ we performed the optimization to calculate the threshold probability $p_{th}^{\mathcal{S}}(n,k)$ for $k=2\ldots n-2$, by which it was found that the optimal number of excitations $k$ is far smaller than $\lfloor \frac{n}{2}\rfloor$, as can be seen in Fig.~\ref{HardyEx100} for the case of $n=100$. We observe that the optimal number of excitations $k$ increases slowly with increasing $n$ (see Fig. \ref{HardyExN}). Unfortunately we were not able to determine their exact relationship.

\begin{figure}[htb]
\begin{center}
\includegraphics[width = \columnwidth]{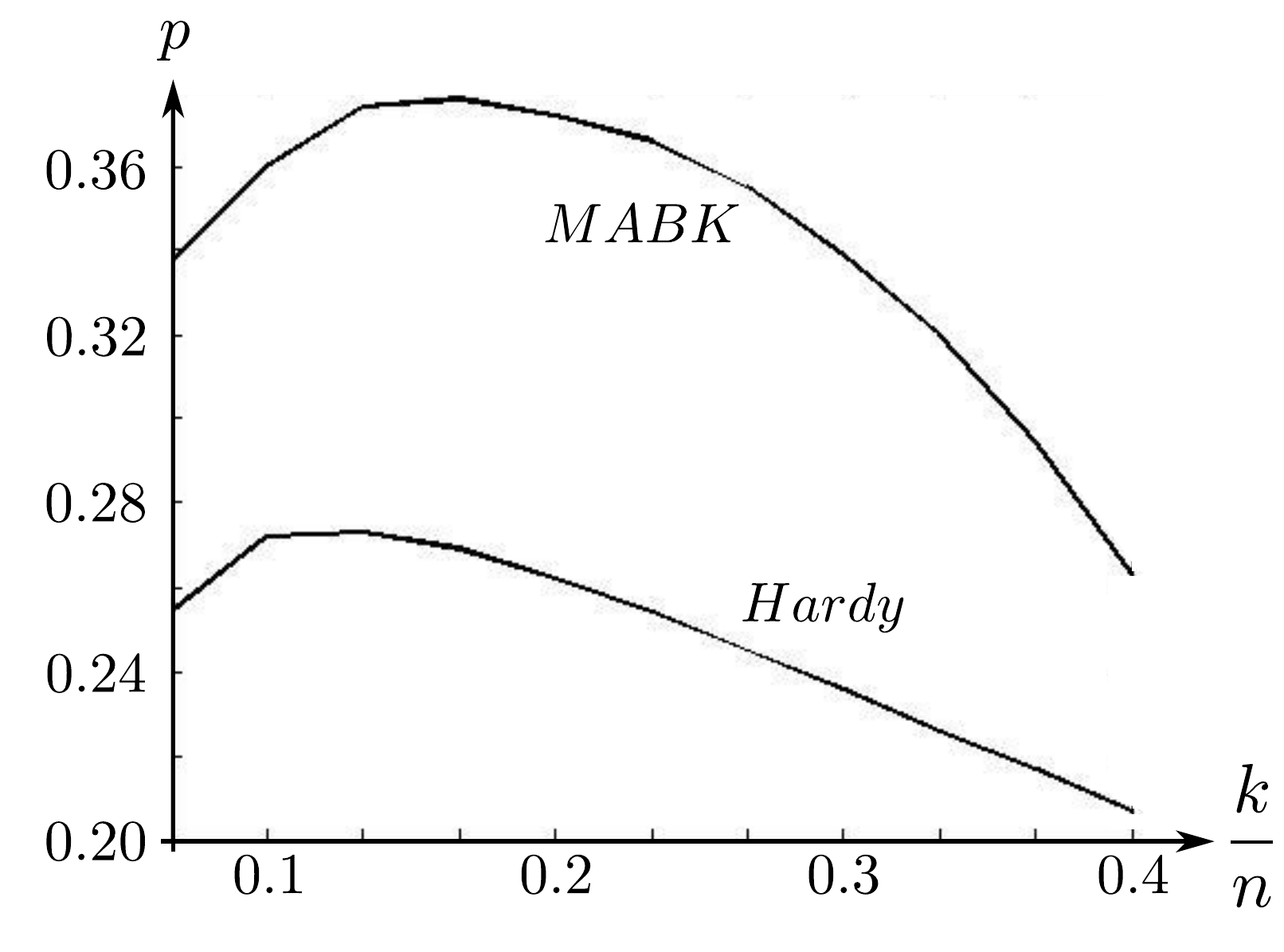}
\caption{A comparison of the threshold values $p_{th}(n,k)$ for the Hardy and MABK inequalities. Here we consider the case $n=30$ qubits. Again the MABK slightly outperform the Hardy inequalities. Interestingly, the most robust states, for each inequality, have only few excitations.}
\label{HMCombEx30}
\end{center}
\end{figure}

The asymmetry of the noise model clearly manifests itself in these findings since we do not observe a symmetry around $k=\lfloor\frac{n}{2}\rfloor$. Nevertheless, the rapid decline of the threshold probability $p_{th}^{\mathcal{S}}(n,k)$ was unexpected and could have been an artifact of our chosen Bell inequality. This prompted us to redo the computations using the MABK inequality (for details see Appendix \ref{C}). The results, which are presented in Fig. \ref{HMCombEx30} for $n=30$, however showed similar behavior, the threshold values $p_{th}^{\mathcal{S}}(n,k)$ and $p_{th}^{\mathcal{M}}(n,k)$ both attain their maximum for a small number of excitations. It can also be seen that the threshold probability given by the MABK inequality is larger than the one given by the Hardy inequality, however the optimization quickly becomes unstable for larger values of $k$ in the MABK case.

We conclude that the most robust state against excitation loss, at least when considering symmetric equatorial measurements, is likely one with only few excitations.

\subsection{Losing particles}
\begin{figure}[htb]
\begin{center}
\includegraphics[width = \columnwidth]{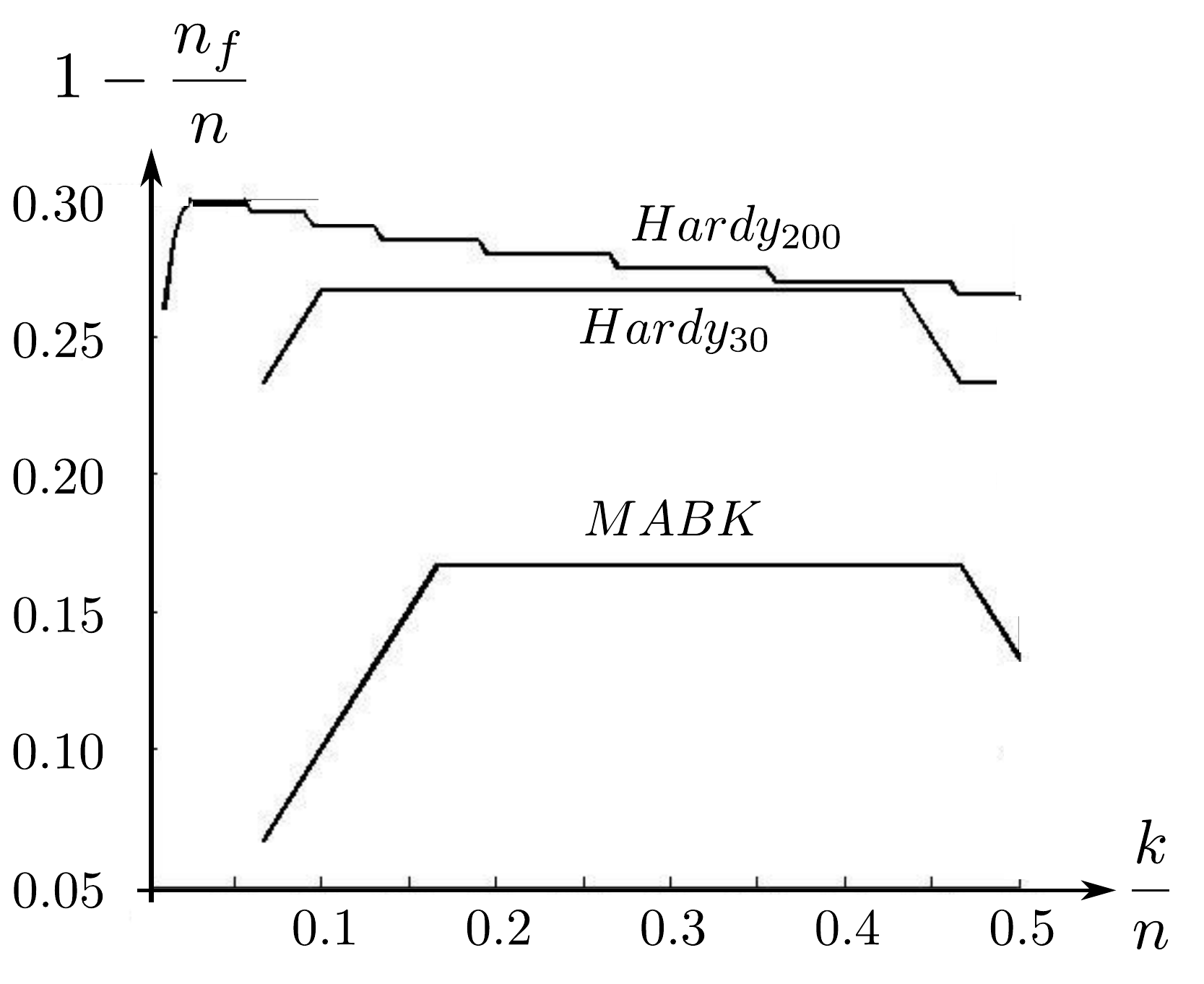}
\caption{Robustness of Dicke states of $n$ qubits with $k$ excitations to loss of particles. We give bounds on the number of qubits in the final state, $n_{f}$, for the Hardy and MABK inequalities. For the Hardy inequality we consider $n=30$ and $n=200$, while for the MABK we restrict ourselves to $n=30$ for computational reasons.}
\label{DHP200}
\end{center}
\end{figure}

The analysis for the case of losing particles is performed in a similar fashion to the case of losing excitations. The main difference is that $\frac{n_f}{n}$ can only take a discrete number of values for fixed $n$. Also, as noted previously, the final state $\tau_f$ is symmetric under $k\rightarrow n-k$ and $\ket{0}\leftrightarrow\ket{1}$, which is why we can limit our analysis to $k\leq \lfloor\frac{n}{2}\rfloor$. We perform the optimization for fixed $n$ and varying values of $k$ for both inequalities. The critical fraction of particles one can afford to lose in order for $\tau_f$ to allow for violations of the specified inequality with symmetric equatorial measurements is shown in Fig. \ref{DHP200}, for the Hardy inequality ($\mathcal{S}_{30}$ and $\mathcal{S}_{200}$) and for the MABK inequality ($\mathcal{M}_{30}$). Interestingly, comparing the inequalities for a fixed value of $n=30$ in this case shows that the Hardy inequality clearly outperforms the MABK one.

As with the case of losing excitations, we notice again that the highest robustness is achieved for small (and in this case by symmetry also large) numbers of excitations, and not around $k=\lfloor\frac{n}{2}\rfloor$. This is further evidence that for the robustness of nonlocal properties, the state with the largest entanglement $k=\lfloor\frac{n}{2}\rfloor$ may not be optimal.\\

\section{Conclusion}

We have investigated the nonlocality of symmetric Dicke states of $n$ qubits subject to losses. We considered two models of losses, namely loss of excitations and loss of particles. For each loss model, we investigated the robustness of the nonlocality of these states using two different families of (multipartite) Bell inequalities. We found that independent of $n$, the most robust Dicke states are those featuring a small number of excitations, i.e. $k\ll n$. Since Dicke states become more entangled when $k$ is close to $n/2$, our results suggest that the relation between nonlocality and entanglement of Dicke states may not be monotonous.

However, this work only marks the beginning of the investigation of the nonlocality of Dicke states when subject to losses. More work will be needed to see whether the behavior observed here is generic, or whether it is due to the fact that we focus on two specific classes of Bell inequalities and the restriction to symmetric equatorial measurements. In particular it would be interesting to consider Bell inequalities with more measurement settings per party and less symmetric measurements. Another interesting aspect would be to study the robustness of genuine multipartite nonlocality for Dicke states. Answers to any of these questions would certainly represent significant progress in our understanding of the nonlocal properties of Dicke states.

\emph{Note added.} While finishing writing up the present manuscript, we became aware of related work by Sohbi {\em et al.}~\cite{sohbi14}. In particular these authors also discuss the robustness of the nonlocality of Dicke states upon losing excitations, using the Hardy inequality.

\begin{acknowledgments}

We thank Jean-Daniel Bancal, Peter Divianszky, Florian Fröwis, Denis Rosset, Tamas Vertesi for discussions. We acknowledge financial support from the Swiss National Science Foundation (grant PP00P2\_138917, NCCR QSIT Director's reserve, SNSF Starting Grant DIAQ), and SEFRI (COST action MP1006).

\end{acknowledgments}

\appendix

\section{Values of $S_n$ and $M_n$ for equatorial measurements}
\label{A}

In the following appendices we denote $\cos(\alpha_j)$ by $c_j$ and $\sin(\alpha_j)$ by $s_j$ to simplify notation. The equations for $\mathcal{S}_n(\rho_{n,1},\alpha_0,\alpha_1)$ and $\mathcal{S}_n(\rho_{n,0},\alpha_0,\alpha_1)$ for the Hardy inequality are given below:

\begin{IEEEeqnarray}{rCl}
\label{hardy_W}
\mathcal{S}_n(\rho_{n,1},\alpha_0,\alpha_1) & = & n c_0^{2(n-1)} s_0^2 - n c_1^2 s_1^{2(n-1)} - \nonumber \\ & & \left(c_0^{n-1} s_1 + (n-1) c_0^{n-2} s_0 c_1 \right)^2  \nonumber \\ \mathcal{S}_n(\rho_{n,0},\alpha_0,\alpha_1) & = & c_0^{2 n} - n c_0^{2(n-1)} c_1^2 - s_1^{2 n}
\end{IEEEeqnarray}

The full correlators that are needed to insert into the expression given in \eqref{MN} for the MABK inequality were derived in Eq. $(13)$ of \cite{brunner}. In doing so closed expressions for $\mathcal{M}_n(\rho_{n,1},\alpha_0,\alpha_1)$ and $\mathcal{M}_n(\rho_{n,0},\alpha_0,\alpha_1)$ were derived.

\begin{IEEEeqnarray}{l}
\label{MABK_W}
\mathcal{M}_n(\rho_{n,1},\alpha_0,\alpha_1) = \frac{\sqrt{2}}{4 c_0^{4} c_1^{4}} (1+i) e^{-i n \frac{\pi}{4}} \\
\left[\left(c_0 + i c_1\right)^2 \left(i c_0+ c_1\right)^n \left(2i \left(c_0 c_1 + s_0 s_1\right) + n \left(s_0 -i s_1\right)^2\right) - \right. \nonumber \\ 
\left. \left(c_0 - i c_1\right)^2 \left(c_0+i c_1\right)^n \left(2 \left(c_0 c_1 + s_0 s_1\right) + i n \left(s_0 + i s_1\right)^2\right) \right] \nonumber 
\end{IEEEeqnarray}

\begin{IEEEeqnarray}{rCl}
\label{MABK_W}
\mathcal{M}_n(\rho_{n,0},\alpha_0,\alpha_1) & = & \frac{1}{2} e^{-i (n+1) \frac{\pi}{4}} \\ & & \left[\left(c_0+i c_1\right)^n + i \left(i c_0+c_1\right)^n \right] \nonumber
\end{IEEEeqnarray}

\section{Optimal measurement angles for the MABK inequality with W states}
\label{B}

Here we provide the ansatz for the measurement angles $\alpha_j^\mathcal{M}$ for $n=0,1,2,3 \text{ mod } 4$. For the case $n=0 \text{ mod } 4$ we have

\ba \alpha^{\mathcal{M}}_{0}\left(n\right) & = & \frac{\pi}{2} + \arctan\left(\frac{5}{4} \sqrt{n}\right) \nonumber \\
\alpha^{\mathcal{M}}_{1}\left(n\right) & = & \frac{\pi}{2} - \arctan\left(\frac{4}{9} \sqrt{n}\right).\ea
For $n=1 \text{ mod } 4$ we have
\ba \alpha^{\mathcal{M}}_{0}\left(n\right) & = & \frac{\pi}{2} + \arctan\left(0.72\sqrt{n}\right) \nonumber \\
\alpha^{\mathcal{M}}_{1}\left(n\right) & = & \frac{\pi}{2} - \arctan\left(\frac{4}{3}\sqrt{n}\right). \ea
For $n=2 \text{ mod } 4$ we have
\ba \alpha^{\mathcal{M}}_{0}\left(n\right) & = & \frac{\pi}{2} - \arctan\left(0.72\sqrt{n}\right) \nonumber \\
\alpha^{\mathcal{M}}_{1}\left(n\right) & = & -\frac{\pi}{2} + \arctan\left(\frac{4}{3}\sqrt{n}\right).\ea
Finally, for $n=3 \text{ mod } 4$ we have
\ba \alpha^{\mathcal{M}}_{0}\left(n\right) & = & \frac{\pi}{2} - \arctan\left(\frac{3}{4}\sqrt{n}\right) \nonumber \\
\alpha^{\mathcal{M}}_{1}\left(n\right) & = & \frac{\pi}{2} + \arctan\left(\frac{4}{3}\sqrt{n}\right).\ea

\section{Probabilities and correlators for symmetric equatorial measurements on Dicke states}
\label{C}

The expressions $\mathcal{S}_n(\rho_{n,k},\alpha_0,\alpha_1)$ for the Hardy inequality are given by a generalization of Eq.~\eqref{hardy_W}. The value $E(\vec{x})$ for the Dicke states can also be found below. Specifically, we have that
\begin{widetext}
\begin{IEEEeqnarray}{rCl}
\label{hardy_Dicke}
\mathcal{S}_n(\rho_{n,k},\alpha_0,\alpha_1) & = & {n \choose k} c_0^{2(n-k)} s_0^{2 k} - \frac{n}{{n \choose k}}\left[{n-1 \choose k-1} c_0^{n-k} s_1 s_0^{k-1}+{n-1 \choose k} c_0^{n-1-k} c_1 s_0^{k} \right]^2 - {n \choose k} s_1^{2(n-k)} c_1^{2 k}
\end{IEEEeqnarray}
\begin{IEEEeqnarray}{rCl}
\label{dicke_pauli}
E(\vec{x}) & = & \frac{1}{{n_f \choose k}} \sum_{r=0}^{k} \sum_{q=\max(0,2r+x-n_f)}^{\min(2r,x)} (-1)^{k-r} {n_f-x\choose 2r-q}{x\choose q}{2r\choose r}{n_f-2r\choose k-r} c_0^{n_f+q-x-2r} s_0^{2r-q} c_1^{x-q} s_1^q.
\end{IEEEeqnarray}
\end{widetext}
Note that, contrary to the case of the $W$ state, a closed expression for $\mathcal{M}_n(\rho_{n,k},\alpha_0,\alpha_1)$ [computed using Eq.~\eqref{MN}] could not be found.

\end{document}